\documentclass[10pt, conference]{IEEEtran}
\usepackage{cite}
\usepackage{xcolor}
\usepackage{amsmath,amssymb,amsfonts}
\usepackage{import}
\usepackage{balance}
\usepackage{braket}
\usepackage{adjustbox}
\usepackage{graphicx}
\usepackage{adjustbox}
\usepackage{soul}
\setstcolor{red}

\def\BibTeX{{\rm B\kern-.05em{\sc i\kern-.025em b}\kern-.08em
    T\kern-.1667em\lower.7ex\hbox{E}\kern-.125emX}}
\begin{document}


\title{Quantum Leak: Timing Side-Channel Attacks on Cloud-Based Quantum Services}  


\author{

\IEEEauthorblockN{Chao Lu$^{\dag}$, Esha Telang$^{\ddag}$, Aydin~Aysu$^{\ddag}$, Kanad Basu$^{\dag}$}
\IEEEauthorblockA{$^{\dag}$Department of Electrical and Computer Engineering, University of Texas at Dallas, TX, USA\\
                $^{\ddag}$Department of Electrical and Computer Engineering, North Carolina State University, NC, USA
}
}

\maketitle

\begin{abstract}

Quantum computing offers significant acceleration capabilities over its classical counterpart in various application domains. Consequently, there has been substantial focus on improving quantum computing capabilities. However, to date, the security implications of these quantum computing platforms have been largely overlooked. With the emergence of cloud-based quantum computing services, it is critical to investigate the extension of classical computer security threats to the realm of quantum computing.

In this study, we investigated timing-based side-channel vulnerabilities within IBM's cloud-based quantum service.
\textcolor{black}{The proposed attack effectively subverts the confidentiality of the executed quantum algorithm, using a more realistic threat model compared to existing approaches.} Our experimental results, conducted using IBM's quantum cloud service, demonstrate that with just 10 measurements, it is possible to identify the underlying quantum computer that executed the circuit. Moreover, when evaluated using the popular Grover circuit, we showcase the ability to leak the quantum oracle with a mere 500 measurements. These findings underline the pressing need to address timing-based vulnerabilities in quantum computing platforms and advocate for enhanced security measures to safeguard sensitive quantum algorithms and data. 
\end{abstract}


\section{Introduction} \label{sec:intro}

Security should not be an afterthought but instead be embedded into the design process from the beginning. This is especially true for quantum computing technologies, as they currently are finding their first adoptions.
Indeed, quantum computing is considered as the next generation of computation, capable of performing tasks that might take an extremely long time for even the most powerful classical computers\cite{quantumsupremacy_nature_2019}. Consequently, substantial efforts have been directed towards enhancing the performance of quantum computation. These efforts encompass various aspects such as fabricating more qubits, improving fidelity, designing and optimizing algorithms, quantum compilation, and verification~\cite{dwave, qiskit, shor, quantumECC}. 
Ongoing improvements in the field of quantum computing including enhancing qubit quality and quantity, as well as the design of novel quantum algorithms, are gradually paving the way for more practical applications of quantum computers. 



Currently, the development and maintenance of an actual quantum computer for practical purposes comes at an extremely high cost. As a result, a majority of computing resources are provided through cloud-based services to mitigate the expenses associated with using quantum computers. Industry giants such as IBM, Google, D-Wave, and Amazon offer cloud-based quantum computation resources to cater to the growing demand~\cite{qiskit, cirq, dwave, amazon}. 
It is imperative to investigate the security of sensitive information when users execute quantum circuits on such quantum cloud services. 
This aspect deserves increased attention from the community in parallel with performance improvement.

Side-channel attacks (SCAs) are a major threat to cyberspace. SCA focuses on analyzing implementation by-products, such as power consumption, timing, or electromagnetic radiation, that inadvertently reveal clues about the inner workings of the system. By meticulously analyzing these side-channel signals, attackers can discern cryptographic keys, passwords, or other confidential data, even if the underlying algorithm or implementation appears to be secure. As a result, SCA poses a significant threat to the security of various computing devices, including smart cards, embedded systems, and cryptographic implementations, necessitating proactive measures to mitigate these covert and often hard-to-detect threats \cite{iot-sca,balasch15,scaapple,genkin14,luo15}.

Prior research in this domain either requires physical access~\cite{ccspscaquantum} or assumes simple quantum circuits with a single parameterized quantum gate~\cite{quantumcloudsca}.
Therefore, it is critical to conduct a comprehensive analysis of remote timing-based vulnerabilities of quantum circuits with fewer restrictions regarding the types and sizes of quantum circuits.

In this paper, we investigated the side-channel security vulnerabilities of a large class of quantum circuits within IBM's quantum cloud service. \textcolor{black}{Compared to~\cite{ccspscaquantum}, the threat model that we utilized requires fewer assumptions and does not need physical access to a quantum computer.} Our findings reveal the successful extraction of sensitive information during the execution of multiple quantum circuits with certain functions. This innovative approach aims to identify instances where other malicious entities might exploit vulnerabilities 
in quantum circuitry and associated properties. This strategy is evaluated through simulations employing quantum simulators as well as data obtained from actual IBM quantum computers hosted on the cloud service. To that end, the threat model that we utilized focuses on the time consumption parsed by IBM's cloud service. Our contributions are listed as follows:

\begin{itemize}
\item We explore the possible security vulnerabilities by performing timing SCAs on IBM's quantum cloud service.
\item We performed five different types of timing SCAs and found that three out of five attacks are vulnerable to the timing SCA that potentially leaks sensitive information. By using statistical tests, we quantified the effort needed to steal such information.
\item We performed a comparative analysis of the proposed cloud-based timing SCAs with the local power-based SCA~\cite{ccspscaquantum}, and evaluated the difference between the two.
\end{itemize}


Our experimental results, on IBM's \textit{ibm\_perth}, \textit{ibmq\_belem} and \textit{ibm\_lagos}, demonstrate that by conducting approximately 10 measurements, we can identify the specific quantum processor in use. Moreover, we can discern the victim's quantum circuit type with a minimum of 1 measurement and a maximum of 18,712 measurements. \textcolor{black}{Furthermore, the popular Grover circuit oracle can be extracted with a minimum of 500 measurements, and a maximum of 20 million timing measurement data from the attacker. }
This work informs cloud providers about potential vulnerabilities and discusses related mitigation strategies for future cloud deployments.

The rest of this paper is organized as follows. 
Section~\ref{sec:background} provides background on SCA, as well as quantum computing basic knowledge and statistic tools for the SCA. Section~\ref{sec:related} provides the related work on the SCAs on classical computers and quantum computers with different approaches. Section~\ref{sec:Methodology} provides the threat model, as well as the procedure for performing the SCA. Section~\ref{sec:Results} presents our experimental results and possible defenses to the proposed timing SCA. Section~\ref{sec:Conclusion} summarizes the paper
\section{Background}\label{sec:background}

This section provides preliminary information on quantum computing technology and statistical tools for SCAs to help with the rest of the paper.

\subsection{Quantum Computing}

\subsubsection{Quantum Computing Basics}
\begin{figure}[bt!]
\centering
\includegraphics[width=\columnwidth]{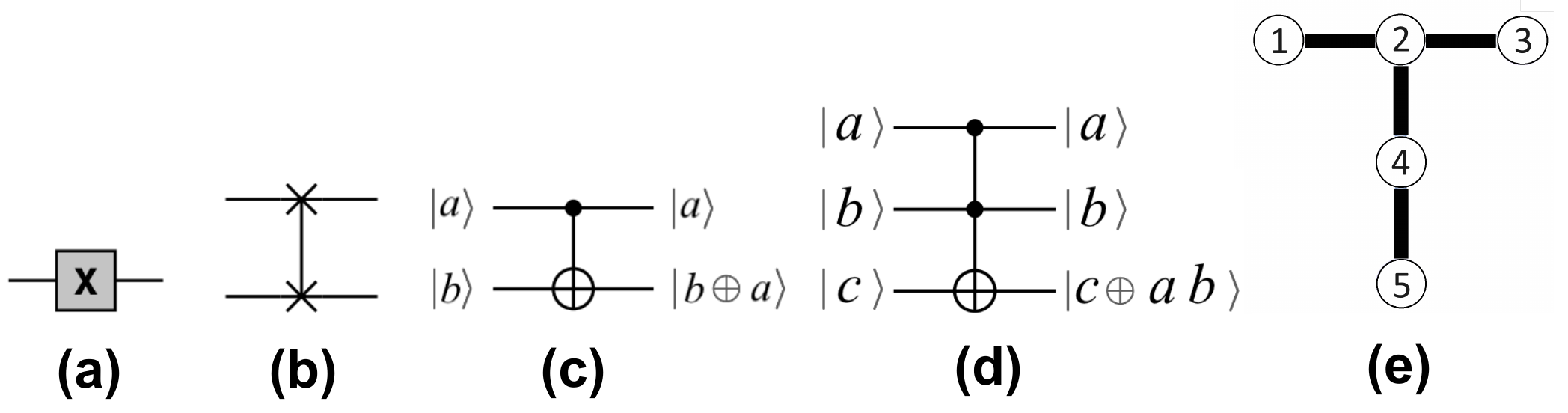}
\caption{(a) X gate (b) CX gate (c) SWAP gate (d) CCX gate. (e) Routing Map of the IBM Quantum Computer \textit{ibmq\_belem}: Only connected qubits can perform 2-qubit gates like the CX gate. If a CX gate is required between two unconnected qubits, SWAP gates are introduced to assist with the computation.}
\label{fig:gates}
\end{figure}

Quantum computing utilizes quantum entanglement and superposition to reshape the field of computation. In this new paradigm, computations are carried out using quantum gates on qubits. Unlike classical computers, quantum computers rely on quantum entanglement and superposition for computation. A quantum gate is represented as a unitary matrix, with a matrix size of $2^n$, where $n$ represents the number of qubits. Figure~\ref{fig:gates} illustrates common quantum gates.
These gates allow qubits to become entangled and enable computations that theoretically exhibit exponential acceleration compared to classical computers across various aspects. An \textit{X gate} (Figure 1 (a)) will flip the state of a qubit, which is similar to an inverter in classical computing; a \textit{SWAP gate} (Figure 1 (b)) will swap the information of two qubits. A \textit{CX gate} (Figure 1 (c)) performs $xor$ logic of $a\ \&\ b$, and replaces the output at register $b$ while qubit a will remain unchanged; A \textit{CCX gate} (Figure 1 (d)) performs exclusive or of $a\ \&\ b$ and gives the output at the qubit $c$.
%

\subsubsection{Quantum Compilation} \label{sec:compilation}
Since real quantum computers only support specific quantum gates in their hardware, it is worth noting that IBM's quantum computer, for example, is limited to the gate library \texttt{\{`RZ', `ID', `SX', `X', `CX'}\}~\cite{qiskit}. To enable the implementation of various quantum gates, a logic synthesis approach is essential, converting each quantum gate into its corresponding basic gate from the predefined library. Additionally, physical qubit connections are subject to restrictions, which necessitate the incorporation of SWAP gates. Figure~\ref{fig:gates}(e) demonstrates IBM's quantum computer \textit{ibm\_belem} that is accessible in IBM quantum cloud service. Only connected qubits can form reliable CX gate connections. When quantum circuits require computation that does not have physical connections, SWAP gates serve to exchange information between two qubits, allowing for the physical realization of two-qubit gates on any two physical qubits with SWAP gates-based quantum routing techniques.

Given these constraints, optimizing the insertion of SWAP gates and refining routing techniques becomes crucial in minimizing the quantum circuit's depth and the overall gate count. \textcolor{black}{Various efforts have been dedicated to enhancing circuit depth and mitigating noise~\cite{abrams2019methods, sarovar2020detecting}.} In the context of Qiskit, which is a quantum circuit design and execution toolkit integrated for IBM's quantum computing ecosystems, several compilation techniques have been integrated~\cite{qiskit}. In this paper, our focus remains directed at investigating the SABRE and Stochastic routing approaches that are embedded in the qiskit library for easier experimental setup~\cite{sabre, qiskit}. Our aim is to explore potential vulnerabilities in the qubit mapping process and the feasibility of attacks on it.


\subsubsection{Quantum computer architecture}

\begin{figure}[bt!]
\centering
\includegraphics[width=\columnwidth]{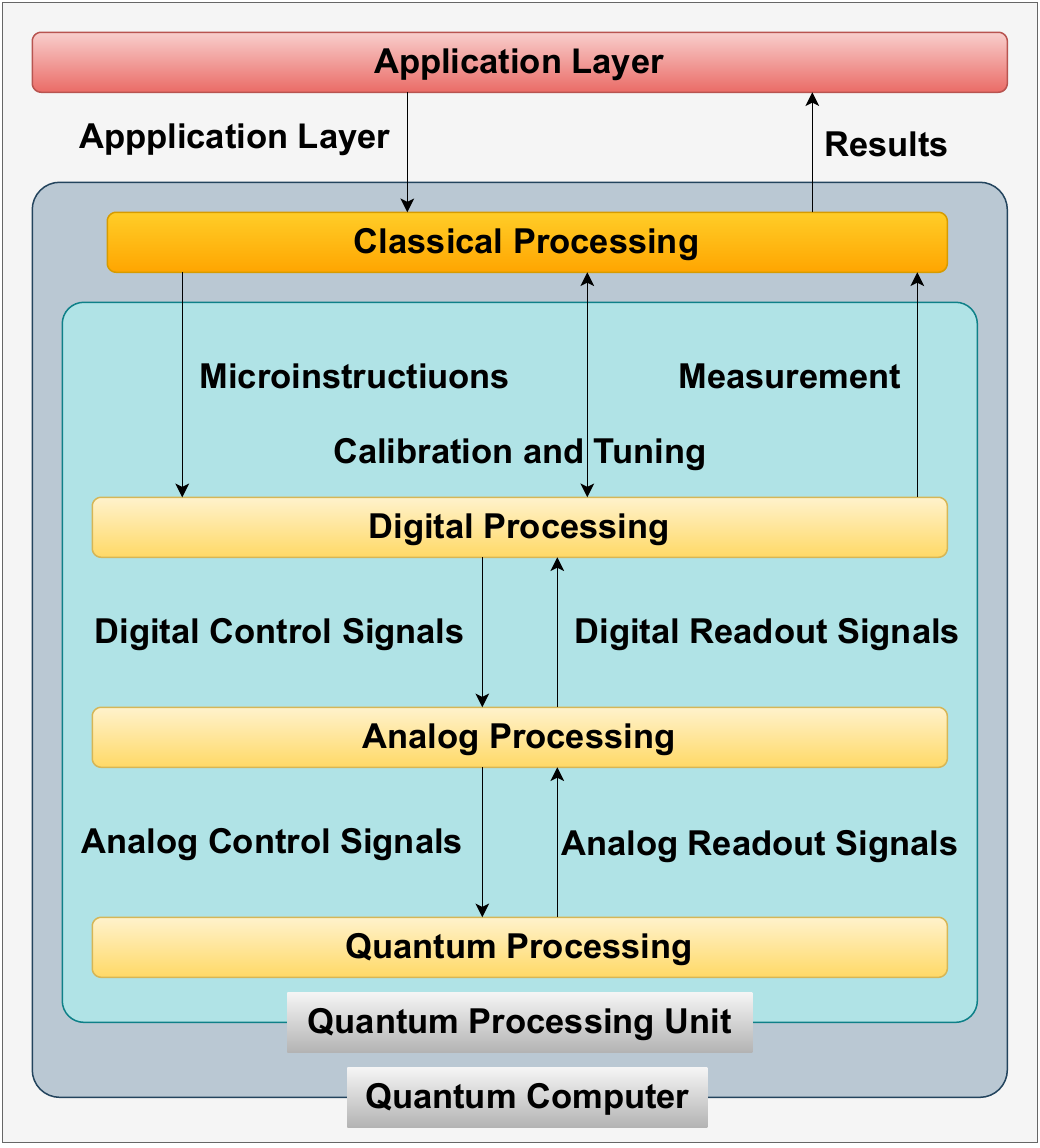}
\caption{Quantum computer execution process: The quantum circuits submitted by users are processed in the application layer, following the workflow shown in the Figure. In this paper, we will only evaluate the time consumption in the Quantum Processing Unit, as the time consumption during calibration and tuning interactions between the Classical Processing section and the Digital Processing section is unstable.}
\label{fig:quantumarch}
\end{figure}

\textcolor{black}{The process of cloud-based quantum computing requires the following steps.} Firstly, the quantum circuit is designed and compiled on the classical computer. This process includes steps such as logic synthesis and layout synthesis. Next, the quantum circuit is submitted into a quantum computer in the cloud. The cloud server will start converting the transpiled circuit into a set of pulse wave instructions. Finally, the quantum computer will convert the digital signals to analog signals and control the quantum computer to generate certain light beams on a superconducting Josephson junction to perform the computation~\cite{josephsonjunctions}.
Once the circuit finishes the computation, the quantum computer will measure the quantum processors and send the results to the user. The overall computation structure is demonstrated in Figure~\ref{fig:quantumarch}.

\subsection{Statistical Analysis for SCAs}

SCAs leverage leaked information from computing by-products such as power consumption, electromagnetic emissions, time measurements, and even sound produced during circuit execution. 
By gathering this data, attackers can extract sensitive information from electronic circuits, and they can steal cryptographic keys from mathematically-secure systems.
These attacks can also steal other information such as the browsed website, ATM pins, or machine learning models~\cite{martin2015counting, martinovic2012feasibility, gupta2023ai, weinberg2011still}.
To this end, SCAs use statistical techniques to extract sensitive information and its correlation to secrets from noise.

Timing SCAs are an important instance of such attacks because they do not need physical access to the device, making them especially important for cloud systems.
In our case, these attacks can utilize the time consumed during the execution of a quantum circuit. 
Through the collection of timing data from computations, it becomes possible to infer the private information of a victim~\cite{liu1992forecasting}. Consequently, conducting such an attack necessitates the use of several statistical analysis methods. 
\textcolor{black}{Specifically, we explore the implementation of the t-test, the difference-of-mean test (DoM test), the analysis of the normal distribution's distances, and power analysis, which are described in subsequent subsections.}

\subsubsection{T-test and Difference-of-Mean Test} \label{sec:ttest}

The t-test is a statistical tool employed to ascertain whether two data sets are drawn from the same population. It operates under the null hypothesis that two groups of data possess different mean values. Inputs to the t-test include the means, standard deviations, and sample sizes of the datasets. The t-test produces a t-value or t-score as the output, indicating whether the datasets differ significantly. Equation~\ref{eq:t-test} illustrates this relationship for the independent two-sample t-test, where $\mu$ represents the mean values, $\sigma$ represents the standard deviations, and $N$ signifies the sample sizes of each data set.
\begin{equation}
\label{eq:t-test}
t = \frac{{\mu}_1 - {\mu}_2}{\sqrt{\frac{{\sigma}_1^2}{N_1}-{\frac{{\sigma}_2^2}{N_2}}}}
\end{equation}


Upon completion of the calculation, the resulting t-score signifies the disparity between the two input datasets. A larger t-score indicates a greater separation between these datasets. 
This mathematical equation serves as a fundamental tool in SCA for distinguishing between two datasets obtained from two different configurations.

The difference-of-mean (DoM) test is an alternative form of the t-test. A DoM test calculates the difference of the means and plots it into a graph. After the computation is finalized, the obtained result can be contrasted against a t-distribution table with a specified confidence level. In such scenarios, a figure can be plotted, allowing for the calculation of the mean values of the two datasets. A t-critical value is also incorporated, accounting for the standard deviation of the two input datasets, determining whether these datasets stem from the same group. The DoM test dynamically discriminates between datasets as the sample size grows, making it easier to observe and determine the dissimilarity between the two datasets with respect to the number of measurements.

\subsubsection{Power Analysis} \label{sec:power analysis}


Power Analysis, not to be confused with power-based SCAs, can be used to determine the minimum sample size required for distinguishing two distinct Gaussian-distributed datasets, given their mean and standard deviation. Additionally, two more parameters are necessary for the calculation: \textit{significance} and \textit{statistical power}, which govern the types of errors that the analysis could encounter. Lehr's approximate guideline suggests that the significance can typically be set at $0.05$, and the statistical power at $80\%$, to generate a dependable outcome for determining the minimum samples needed to differentiate two independent datasets~\cite{lehr1992sixteen}. If the overlapping coefficient of the two Gaussian-distributed samples is close, a greater number of samples is usually required to distinguish the two datasets.

\begin{figure}[bt!]
\centering
\includegraphics[width=\columnwidth]{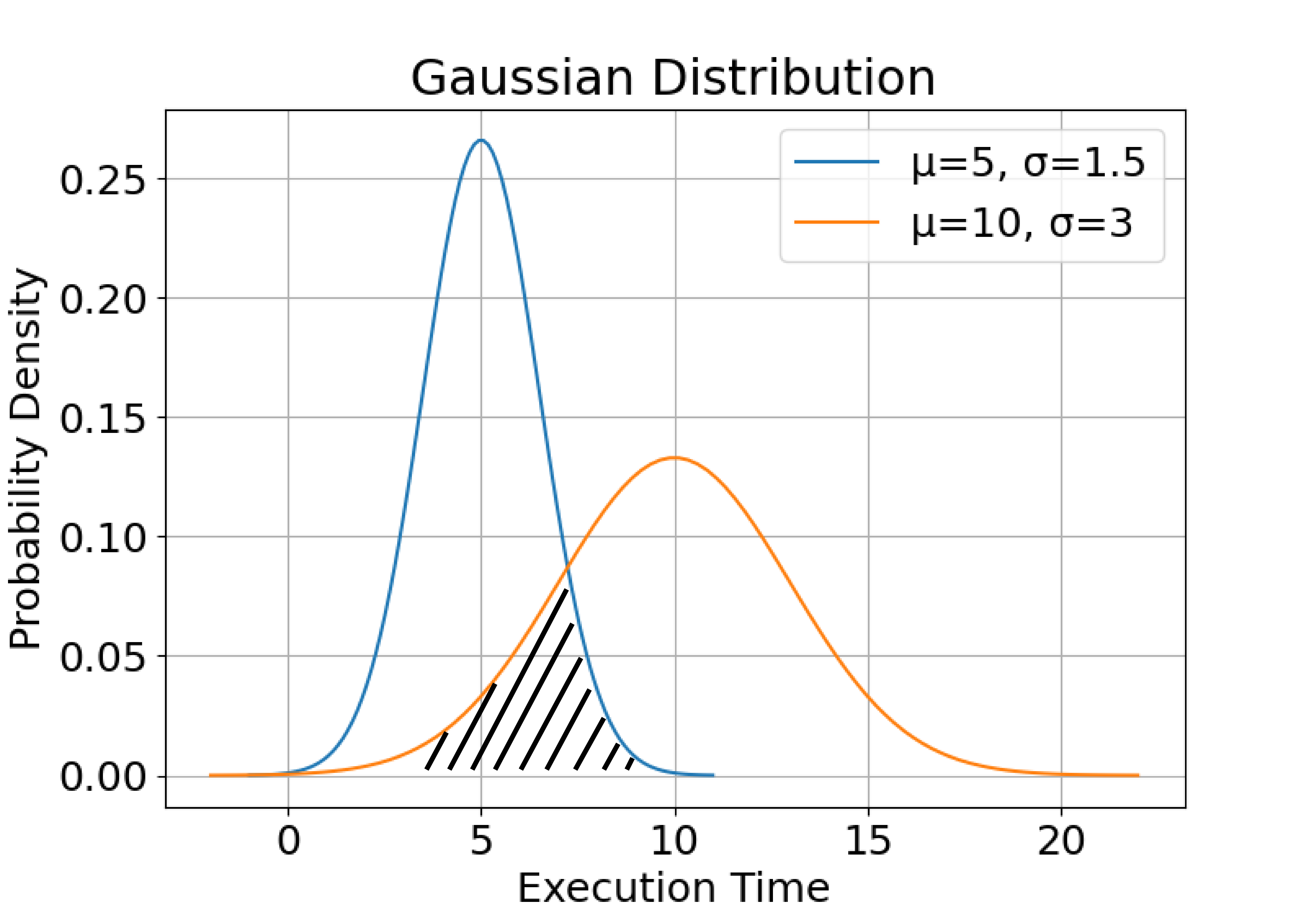}
\caption{Demonstration of the overlapping coefficient of two Gaussian distributed curves. The shaded area represents the overlapping coefficient of the two curves. When the overlapping coefficient of the two curves is higher, it indicates that the two curves are more similar to each other. 
}
\label{fig:normaldis}
\end{figure}

The overlapping coefficient signifies the extent of overlap between two Gaussian-distributed curves. It quantifies the distance between the areas covered by the two Gaussian distributions, as depicted by the shaded region in Figure~\ref{fig:normaldis}. From the figure, the ratio of the shaded area to the total covered area of the two Gaussian curves along the $x$-axis represents the divergence between the two datasets. If the ratio of these shaded areas is high, a larger sample size is needed to distinguish the two distinct datasets. Thus, it reflects the similarity between the two datasets.

\section{Related Works} \label{sec:related}
SCAs are typically performed on modern CMOS-based systems, micro-controllers in IoT/embedded applications~\cite{iot-sca}, modern processors~\cite{balasch15}, smartphones~\cite{scaapple}, multi-core CPUs~\cite{genkin14}, and GPUs~\cite{luo15}. Such attacks can extract secret keys, steal pincodes from ATMs, learn the details of a trained AI/ML model, and even identify the website being browsed~\cite{martin2015counting, martinovic2012feasibility, gupta2023ai, weinberg2011still}. These attacks utilize computation byproducts, such as execution time, power consumption, or electromagnetic emissions.  


Recent work has defined a taxonomy of SCA on quantum computers and used power consumption information to extract sensitive information~\cite{ccspscaquantum}\footnote{Note that the paper also performs a timing attack for one of the scenarios; but even for that scenario, the adversary is assumed to have physical access to measure the control pulses and derive execution time from that information~\cite{ccspscaquantum}.}. Specifically, the paper assumes that the adversary can capture the impact of radio frequency control pulses that execute quantum gate operations in single or two-qubit pairs. However, such attacks have low viability, since they require an insider to record the power traces of quantum computers. It might be easier to obtain the quantum gate instructions from the digital and analog processes to obtain the user's quantum circuits. Moreover, quantum computers are extremely expensive to build and maintain, quantum cloud service is the main methodology that allows users to get access. Therefore, this type of attack is infeasible in a cloud setting where the adversary does not have physical access to the quantum computer.

Another study has delved into the vulnerability of quantum cloud services, employing SCAs to extract the hidden parameter in the victim circuit~\cite{quantumcloudsca}. This involved executing a dummy circuit before and after the execution of the victim circuit and extracting the dummy circuits' readout information to estimate the angle of the victim quantum gate. Such attacks can be performed on IBM's quantum cloud service without direct access to the quantum physical hardware. However, this approach faces limitations when dealing with quantum circuits containing multiple gates or quantum gates without any parameters. Furthermore, a single parameterized gate as a quantum circuit is insensitive to the victim. 
\textbf{By contrast, our novel cloud-based quantum timing SCA overcomes both the physical access requirement and the quantum circuit constraints, enabling the extraction of information from the victim's quantum circuits execution with less constraint.}

In the realm of quantum computing security, substantial research has been devoted to safeguarding the execution of quantum circuits, particularly in the context of 
multi-programming systems~\cite{ash2020analysis, deshpande2022towards,jakup-quantum1}. Based on the specification, a potential attack strategy within this domain necessitates the concurrent execution of multiple programs, allowing them to run without any limitations on parallelism. Another quantum security algorithm adopts an alternative approach, employing reset functions within a spacial multi-tenant system to estimate rotational angles of the quantum setup~\cite{jakup-quantum1}. This security approach capitalizes on the inherent imperfections of qubit physical properties to gain access to victim circuits. The threat model then manipulates the original outcomes, rendering the attack easily executable by the victim. 

Furthermore, quantum cross-talk serves as a method to compromise victim circuits without introducing new quantum gates to the original configuration~\cite{ash2020analysis}. This approach utilizes a multitude of CX gates to acquire the qubit state of the target quantum bit~\cite{deshpande2022towards}. It is crucial to note that this attack could only successfully extract the state of the specific target qubit, as opposed to compromising the entire quantum circuit. 
These prior attacks require a multi-programming system where multiple quantum circuits are executed at the same time in the quantum system. Therefore, the attackers have the authority to modify the victim's quantum circuit or append a redundancy in the quantum computer to achieve the attack goal. By contrast, our proposed timing SCA does not require a multi-programming system and the privileged accessibility of the quantum circuit execution to obtain sensitive information.
\section{Threat Model} \label{sec:Methodology}
In this section, we will delve into the details of the attacks we have executed. Specifically, we aim to elucidate the threat model that underpins the general methodology guiding this paper's composition. 

\begin{figure}[bt!]
\centering
\includegraphics[width=\columnwidth]{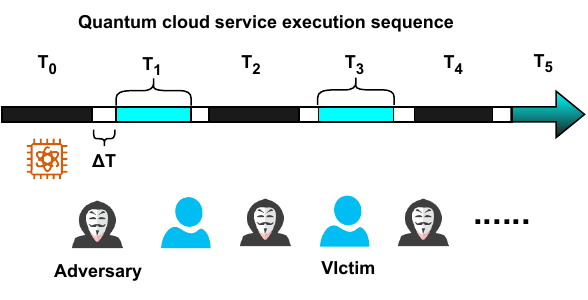}
\caption{Demonstration of the threat model that combining the advantage of~\cite{ccspscaquantum} and~\cite{quantumcloudsca}. The attacker measures the time interval between two circuit submissions to the cloud service and uses a statistical analysis tool to perform SCAs. Our analysis assumes that the time consumed within these intervals is either neglected or treated as a constant. This assumption ensures that the variable $\Delta T_n$, as referenced in the Figure, does not impact the statistical characterization of the execution of diverse quantum circuits.}
\label{fig:threatmodel}
\end{figure}

\textcolor{black}{To achieve a more realistic attack compared to existing research, we are using the attack taxonomy of~\cite{ccspscaquantum} and combining it with the realistic threat vector outlined in~\cite{quantumcloudsca}. Our threat model performs SCA on the cloud while trying to achieve similar attack goals proposed in~\cite{ccspscaquantum}. However, unlike ~\cite{ccspscaquantum}, our proposed attack does not require physical access to the quantum computer.}
Figure~\ref{fig:threatmodel} outlines the side-channel threat model~\cite{quantumcloudsca}. There are three parties involved in our scenario: quantum cloud services, the victim user running a quantum circuit, and the adversary. The victim circuit can belong to either the quantum cloud service or a third-party IP provider. The adversary circuit is executed between the victim circuit's execution phases.
From the figure, the adversary can calculate the execution time of the victim circuit by calculating the difference between the start time of the second job from the end time of the first job. We assume that this calculated time interval closely approximates the actual execution time of the victim circuit. This aims to retrieve time consumption data, calculate the time consumption of the victim circuit on each quantum circuit, perform statistical analysis, and recover sensitive information based on the retrieved time information.

This paper adopts a comparable threat model to~\cite{ccspscaquantum, quantumcloudsca}.
\textcolor{black}{In this work, we utilize weaker assumptions compared to the prior work, thus, making it more realistic, and reducing the capabilities of the attacker.} Specifically, unlike \cite{ccspscaquantum}, we do not assume the attacker has physical access to power measurement~\cite{ccspscaquantum, quantumcloudsca}. Our assumptions of the attacker's available resources are listed as follows:
\begin{itemize}
\item \textcolor{black}{The victim's circuit with the same operations is executed multiple times, similar to~\cite{ccspscaquantum,quantumcloudsca}.}
\item The attack can measure timing for each execution of a circuit~\cite{ccspscaquantum}. Note this is a more realistic assumption than in~\cite{ccspscaquantum} where the adversary can have physical access to measure power.
\item The attacker circuit can be executed during two executions of the two victim circuit execuitons~\cite{quantumcloudsca}.
\item \textcolor{black}{The attacker does not require power, or energy properties for each execution of a circuit, nor access to the qubit drive equipment, where the attacker can collect power and energy traces from the arbitrary waveform generator or the mixer, as was assumed in~\cite{ccspscaquantum}.} 
\item We assume that consecutive executions do not affect the overall time consumption of the quantum circuits, even though the system might take longer to cool down.
\item The randomness of the compilation strategy should be eliminated, meaning that with each compilation of the quantum circuit, the qubit mapping and routing path remain fixed, which is a hidden assumption in~\cite{ccspscaquantum}.
\item The timing SCA encompasses the entire pulse wave preparation stage, comprising digital processing, analog processing, and quantum processing stages, as depicted in Figure~\ref{fig:quantumarch} as the quantum processing unit. The omission of classical processing is due to the relatively shorter time consumption in each computation cycle compared with the other portions of the computation.
\end{itemize}

Note that in cases where multiple victim circuit executions occur between two attacker circuit executions, the attacker can estimate the equal time consumption for each victim execution by comparing it with the average time consumption for the victim's circuit. For example, if the average time consumption for the measured victim's circuit is 3 seconds and the time difference between two quantum circuit measurements from the attacker is 9.1 seconds, the attacker can infer that there were approximately three executions of the victim's circuit between the two measurements of the attacker circuits.

\textit{We use the threat model which only considers threats specifically about quantum computing, not general hardware security risks associated with the hardware infrastructure that handshakes with the quantum computer.}
Other quantum computer security risks involve insecure reset operation to erase leftover data~\cite{jakup-quantum1}, and biases in random number generation~\cite{ghosh-quantum-random}---these are \emph{excluded} in the threat model utilized in the paper and could be addressed by orthogonal defenses.
Unlike prior works, whose threat models are composed of an untrusted server executing quantum computing as a service \cite{swaroop-quantum1, blind-quantum, 9420762, DBLP:conf/iccad/AcharyaS20}, our proposed research assumes the cloud infrastructure itself is trustworthy. We only assume an external attacker in a shared cloud environment who does not collude with the cloud services.
We also note that SCAs on post-quantum cryptography are \textit{irrelevant} in our context, since they analyze classical computers, not quantum.

We assume the attacker is an external party that acquires timing data by conducting attacks between two successive submissions. Through this approach, no malicious insider or victim-related information is necessary to gather the required timing data for computations. However, it is essential to know that the victim user is running the same quantum circuit to construct a dataset, where each element is synchronized throughout the computation. Consequently, an external attacker can execute the attack by submitting a quantum circuit precisely during the overlap of the victim's two quantum circuit submissions. This allows them to obtain the execution time of the victim's quantum circuit.

Moreover, we believe that the major difference in the time consumption is during the quantum processor phase. The other classical elements have minor differences that can be considered as constant time consumption, which are minimal compared with the quantum circuit execution. 
Neglecting the interval between the executions of two quantum circuits allows us to create a timing trace for different executions without considering the time elapsed between them. In cases where the time consumption varies between two executions but remains consistent within each run, recovering the initial timing data becomes seamless. This is accomplished by subtracting the total time consumption during two consecutive executions.

We evaluate five different attacks as categorized in the previous work~\cite{ccspscaquantum}. We summarize the attacks as follows.

\begin{itemize}
\item \textbf{User Circuit Identification (UC)}: This attack is used to identify the quantum circuit being executed in the cloud.
\item \textbf{Circuit Oracle Identification (CO)}: This attack is used to identify the specific quantum oracle utilized.
\item \textbf{Circuits Ansatz Identification (CA)}: Given the certain shape of quantum ansatz, this attack identifies what parameter is utilized in the circuit ansatz. 
\item \textbf{Qubit Mapping Identification (QM)}: Given the known quantum circuit, this attack identifies the qubit routing technique that has been used. 
\item \textbf{Quantum Processor Identification (QP)}: For the given quantum circuit, this attack identifies the underlying quantum hardware.
\end{itemize} 

Note that the reconstruction from power traces mentioned in~\cite{ccspscaquantum} is omitted in this paper. This is because the reconstruction from power trace requires physical access to the quantum computer which the cloud service does not provide. Any attacks that require information on the power traces are inapplicable in the cloud-based timing threat model.

\section{Results} \label{sec:Results}

\begin{table*}[thb!]
\centering
\caption{Time consumption and the minimum number of measurements required to distinguish the quantum circuits of 26 benchmark circuits were evaluated on a Windows 11 desktop using Intel's Xeon W-1165 CPU and the quantum computer \textit{ibmq\_belem}. The latency is represented in seconds.} 
\label{tab:26bench}
\scalebox{0.95}{
\begin{tabular}{lcccc}
\hline
\multicolumn{1}{c}{\textbf{Benchmark Cricuits}} & \multicolumn{1}{c}{\textbf{Simulator Latency}} & \multicolumn{1}{c}{\textbf{QC Latency}} & \multicolumn{1}{c}{\begin{tabular}[c]{@{}c@{}}\textbf{Measurements required} \\ \textbf{to distinguish with other} \\ \textbf{circuits on simulator}\end{tabular}} & \multicolumn{1}{c}{\begin{tabular}[c]{@{}c@{}}\textbf{Measurements required} \\ \textbf{to distinguish with other} \\ \textbf{circuits on \textit{ibm\_belem}}\end{tabular}} \\ \hline
BB84 and Other   Communication Protocols as Benchmarks     & 0.187888861                & 1.649005752         & 22627.76696                          & 8052.789788                             \\
Bernstein-Vazirani   Algorithm                             & 0.308389425                & 2.624919482         & 327.2420257                          & 198.5609927                             \\
Circuit Layer   Operations Per Second (CLOPS)              & 7.145804826                & 111.2657854         & 9.020466364                          & 8.469689255                             \\
Deutsch-Jozsa   algorithm                                  & 0.007853746                & 1.624821556         & 16.91969842                          & 8052.789788                             \\
Entanglement   of Observable                               & 0.087698759                & 1.39879956          & 6696.34903                           & 2312.767812                             \\
Flexible   Representation of Quantum Images (FRQI)         & 0.160657167                & 2.140655833         & 10049.57668                          & 1789.93328                              \\
GHZ                                                        & 0.168084145                & 2.779299043         & 2495.773047                          & 198.5609927                             \\
Grover   Search Algorithm Benchmark                      & 0.18644619                 & 1.853176702         & 188580.9963                          & 18711.59191                             \\
Hidden Shift   Application Benchmark                       & 0.163739443                & 2.476932629         & 4957.915673                          & 176.1900623                             \\
Quantum Edge   Detection                                   & 0.296375513                & 6.591375877         & 327.2420257                          & 2.773546261                             \\
Quantum Error   Correction Threshold                       & 0.158492327                & 1.747846944         & 9073.426949                          & 483.0009561                             \\
Quantum Phase   Estimation                                 & 0.156213999                & 2.866364875         & 9073.426949                          & 622.2060381                             \\
Quantum Random   Number Generation                         & 0.090350866                & 1.590360014         & 6696.34903                           & 3966.373858                             \\
Quantum   Randomized Cryptography Benchmark                & 0.173368216                & 3.075851148         & 1687.593845                          & 108.2803954                             \\
Quantum State   Tomography                                 & 0.855997801                & 11.52563865         & 5.670414924                          & 1                                       \\
Quantum Volume                                             & 7.068844795                & 110.4676444         & 9.020466364                          & 8.469689255                             \\
Qubit   Spectroscopy                                       & 0.571486712                & 4.76002327          & 7.392247531                          & 2.996233403                             \\
Rabi   Oscillations                                        & 0.989506245                & 36.79551248         & 3.881951184                          & 1                                       \\
Randomized   Benchmarking                                  & 1.349968672                & 48.56663332         & 13.41547536                          & 1                                       \\
Shor's   Algorithm                                         & 0.27088356                 & 3.766080118         & 73.44171152                          & 108.2803954                             \\
T1/Qubit   Lifetimes                                       & 1.288301706                & 75.10748723         & 13.41547536                          & 1                                       \\
T2/Decoherence                                             & 0.75401926                 & 32.88892201         & 6.228606762                          & 1                                       \\
The HHL   algorithm                                        & 0.185946465                & 2.191962891         & 188580.9963                          & 1789.93328                              \\
The Vaidman   Detection Test: Interaction Free Measurement & 0.657969475                & 1.869041506         & 7.392247531                          & 18711.59191                             \\
Tphi Dephase   Benchmark                                   & 3.941862106                & 123.7405151         & 1                                    & 1                                       \\
Web Interface   Approx. Execution Time                     & 0.062270641                & 1.443933489         & 16.91969842                          & 2312.767812                             \\ \hline
\end{tabular}
}
\end{table*}
This section describes our experimental setup and provides the results of different attacks that we performed.

\subsection{Experimental Setup} \label{sec:setup}
We conducted a comprehensive analysis involving both quantum simulation and quantum hardware. Initially, we simulated quantum circuits in order to obtain the time consumption of the simulators. Subsequently, we directed our attention towards IBM's quantum computers, including \textit{ibm\_perth}, \textit{ibm\_lagos} and \textit{ibmq\_belem}. Similar experiments can be conducted on other quantum computers as long as the timing information can be retrieved from the cloud service.

To measure the simulation time, we use Intel's Xeon W-1165 CPU, operating on the Windows 11 platform. 
Our quantum simulation endeavors were realized using Qiskit version 0.44. We also utilized the statistical power analysis tool integrated as a library of python, \textit{statsmodels}, to facilitate power analysis calculations with different configurations. 

\subsection{Experimental Results for the Evaluated Attacks}
We conducted an experimental study aimed at investigating potential security vulnerabilities in quantum computers through timing SCAs. This exploration encompassed various attack models, namely UC in Section~\ref{sec:uc}, CO in Section~\ref{sec:co}, CA in Section~\ref{sec:ca}, QM in Section~\ref{sec:qm}, and QP in section~\ref{sec:qp}. To accomplish this, we systematically evaluated the timing SCA vulnerabilities within each attack model and obtained the correlation between the time consumption and circuit specification. The subsequent subsections provide a comprehensive exposition of the distinct algorithms employed, thereby offering detailed insights into their respective implications. We collected 26 different benchmark circuits that either have a quantum advantage over their classical counterpart or are used for evaluating the performance of quantum circuits. The various benchmark circuits used for our experiments are shown in Table~\ref{tab:26bench} (first column). 

\subsubsection{User Circuit Identification (UC)} \label{sec:uc}
To discern the user's circuit, preliminary data must be collected to establish a baseline for various types of quantum circuits. It is necessary to measure the time consumption across different types of quantum circuits in order to construct this baseline. Once this foundational timing information for executing victim circuits is established, a comparison is made against the execution times of quantum circuits using the established baseline. This comparative analysis results in a schedule of execution times, facilitating the application of diverse tests to quantify the time consumption of each quantum circuit.

In this study, we have curated a comprehensive database that includes simulations of 26 distinct quantum circuits. 
We executed quantum circuit attacks on these diverse quantum circuits using IBM's quantum computer \textit{ibm\_belem}, as outlined in Table~\ref{tab:26bench}. The first column of the table provides the names of the quantum circuits, the second column displays the time consumption for executing each quantum circuit on a simulator, and the third column showcases the average time consumption on IBM's quantum computer, \textit{ibm\_belem}. It should be noted that a similar analysis can also be conducted for other quantum computer configurations. Based on the average measurement, we can obtain a variance of $0.003$ in time consumption for a quantum simulator and a variance of $0.3$ for the real quantum computer.

We conducted power analysis, mentioned in Section~\ref{sec:power analysis}, on the benchmark circuits to estimate the minimum number of measurements required to differentiate between two benchmark circuits using timing information. Once the timing data collection for the victim circuit reaches the minimum required measurements to distinguish the closest benchmark circuits (\textit{e.g.}, the average time consumption for the benchmark circuits \textit{Hidden Shift Application} and \textit{GHZ} are $0.168084145$ seconds and $0.163739443$ seconds, respectively), \textcolor{black}{by performing the power analysis, the minimum sample size required to differentiate between the two datasets is found to be $4958$ measurements. We can also observe that the minimum number of measurements required to distinguish \textit{Hidden Shift Application} and \textit{GHZ} quantum circuits is $188,580$.}
Since we assume a fixed variance for all timing datasets, the difference in means can be interpreted as the proximity of different timing datasets. By comparing the mean of the victim timing datasets with the values in the table, we can identify the type of quantum circuits.

We also emphasize that, based on our benchmark results, it is possible to identify if the computation is actually run on a real quantum computer instead of a simulator. The shortest time difference between the actual vs. simulator timing is $18,712$ vs. $188,581$, which is for the case of the quantum computer \textit{ibmq\_belem}. This difference is significantly smaller than the variance.

Prior research conducted a similar SCA employing power traces to recover the victim's quantum circuits~\cite{ccspscaquantum}. The authors concluded that utilizing the timing trace has a more adverse effect than using the power trace. However, in our proposed timing attack, leveraging the timing trace can effectively recover the victim's quantum circuit with a 100\% success rate, performing equally well as the power trace on the UC attack. This methodology can scaled for other quantum circuits as long as the baseline time consumption of quantum circuits is recorded by the attacker. Considering that obtaining the timing trace is simpler, our proposed timing SCA offers greater practicality compared to power-based SCAs. Here, we attempt to distinguish between types of quantum circuits by applying both the Qiskit simulator and IBM's quantum cloud service. By executing the timing SCA, we can effortlessly obtain the user's circuit by conducting a minimum of 1 measurement and a maximum of $18,172$ measurements to distinguish between any two quantum circuits listed in Table~\ref{tab:26bench}. 



\subsubsection{Circuit Oracle Identification (CO)} \label{sec:co}

\begin{figure}[bt!]
\centering
\includegraphics[width=\columnwidth]{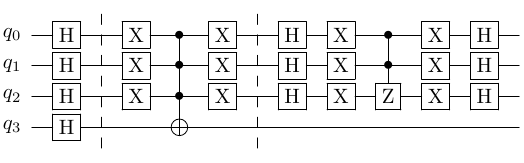}
\caption{Grover search circuit with one iteration. The circuit demonstrates three different parts and separated by dashed lines, which are qubits preparation, Grover oracle, and the diffusion operation. More iterations are done by repeating the Grover oracle and the diffusion operation.}
\label{fig:grover}
\end{figure}

The standard Grover search algorithm is a quantum algorithm that has exponential acceleration on a conventional search algorithm~\cite{grover}. It is composed of three parts, qubits state preparation, Grover oracle, and the diffusion operation, as shown in Figure~\ref{fig:grover}. 
First, the algorithm will prepare the qubits in the superposition state with Hadamard gates. Next, the Grover oracle is appended to the quantum circuit to perform the computation. Then, the diffusion operation will be performed to magnify the desired output. 
If the hidden key is not shown clearly in the result, the Grover circuit can iterate multiple times, so that the computation of the hidden key is easier to obtain. Usually, there is an optimal number of iterations that could maximize the difference between the hidden key and other elements. In this subsection, we consider both the iteration times and Grover oracle for different hidden keys in the CO attack.

\begin{figure}[bt!]
\centering
\includegraphics[width=\columnwidth]{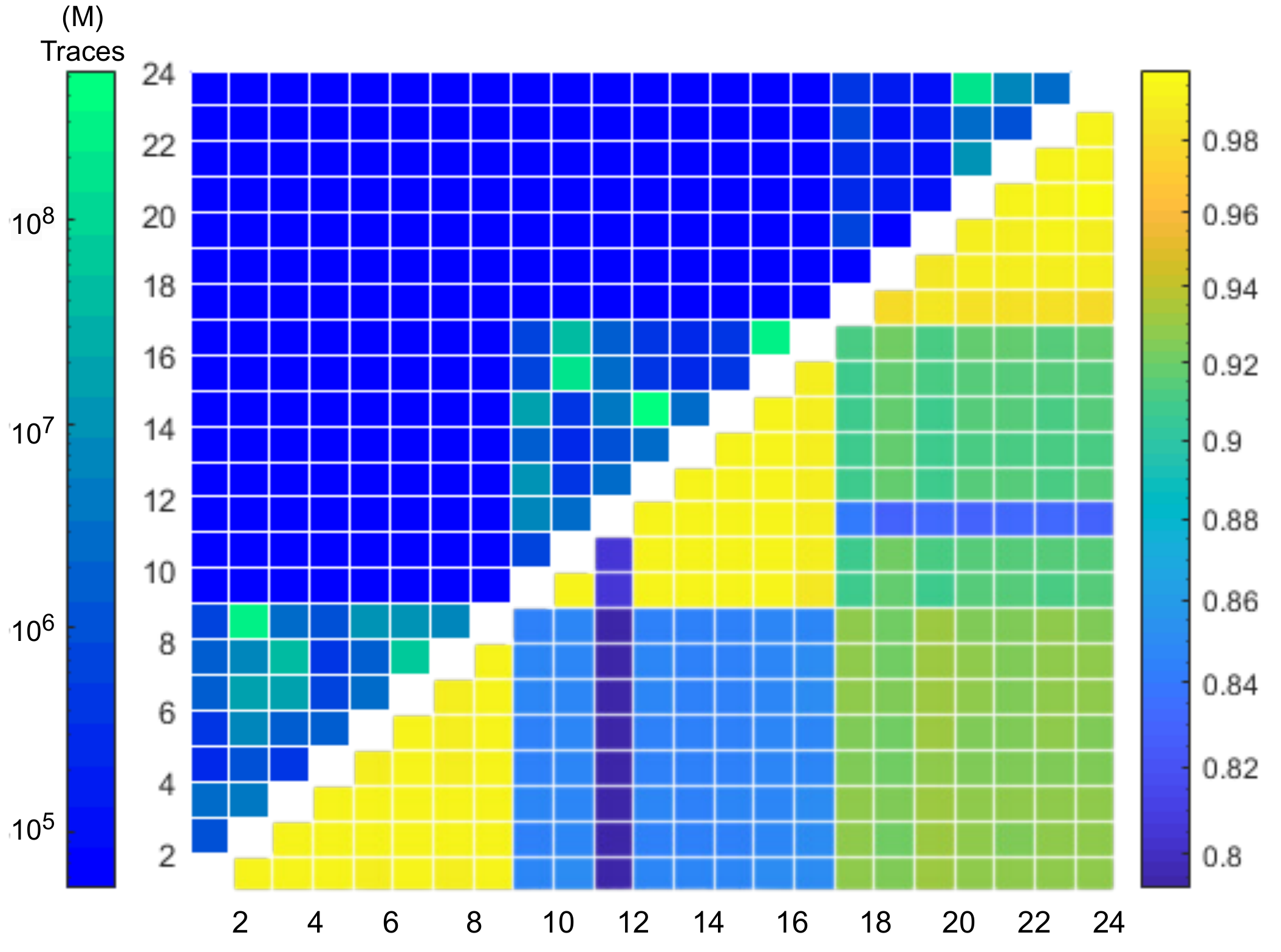}
\caption{Overlapping Coefficient color map (right) and Power analysis (left) on estimating the number of measurements required to distinguish the two Grover oracle executions. It is possible to distinguish the correct circuit in a pairing between 500 to 20M tests.}
\label{fig:groverpower}
\vspace{-5mm}
\end{figure}

This subsection reveals the outcomes of the quantum circuit oracle identification process. The primary aim of quantum circuit oracle identification is to differentiate specific quantum oracles through timing SCA. Our investigation involved gathering time consumption data for the 3-qubit Grover circuit, spanning from one iteration to three iterations. With the 3-qubit configuration yielding $2^3=8$ distinct hidden keys in binary, ranging from $000$ to $111$, and considering three different iteration numbers, the Grover circuit encompasses a total of 24 potential combinations. We obtained the mean and standard deviation metrics from the collected datasets and used the obtained information to generate the overlapping coefficient in Figure~\ref{fig:groverpower}, as explained in Section~\ref{sec:ttest}.
This visualization is generated through a color-coded mapping that illustrates the overlapping coefficients calculated from any two combinations of the 24 distinct datasets, obtained by computing their means and deviations.

Close examination of Figure~\ref{fig:groverpower} unveils the challenge of distinguishing between the two datasets solely based on their overlapping coefficients. The numerical values on the x-axis and y-axis correspond to different types of Grover oracles. Specifically, numbers 1 to 8 pertain to single-iteration circuits with varying quantum oracles corresponding to the 8 distinct hidden keys. Numbers 9 to 16 represent 8 different Grover oracles, each associated with a unique hidden key. Numbers 17 to 24 correspond to 3 iterations of circuits with 8 different oracles. In such instances, we conducted the overlapping coefficient test on any two pairs of Grover oracle and iterations to determine the original Grover oracle. From the figure, it is evident that the overlapping coefficient is higher when the two Grover circuits have the same number of iterations, whereas the coefficient is lower when the iteration numbers are different. This observation implies that, as expected, more measurements are required to distinguish between two datasets when there is minimal difference in the quantum circuits due to the high similarity of quantum oracles.

The obtained results reveal that this challenge arises as most values exceed 0.8, indicating an overlap of over 80\% between the distributions of the two datasets. Power analysis, outlined in Section~\ref{sec:power analysis}, becomes imperative for estimating the necessary sample size to effectively differentiate between the datasets. This sample size directly corresponds to the number of quantum circuit executions required to distinguish the two distinct sets of data. Note that Figure~\ref{fig:groverpower} illustrates the requisite sample size for efficient differentiation between the two samples. As previously mentioned, the $x$- and \textit{y-axes} in the figure denote different Grover circuits.

The results depicted in the figure unveil that distinguishing between two quantum circuits becomes notably more straightforward when they possess differing numbers of iterations. This observation is evident from the blue intersections between Grover circuit numbers 1 to 8 and Grover circuit numbers 9 to 24, as well as between Grover circuit numbers 9 to 16 and Grover circuit numbers 17 to 24. This blue color indicates that only a few hundred measurements are needed to differentiate the two Grover oracles with a different number of iterations. Conversely, when the number of iterations is the same for the respective Grover circuits, a larger number of measurements, likely exceeding a million, is necessary to exploit the victim quantum circuit oracle. Consequently, our proposed statistical results can successfully distinguish any two quantum oracles from the two Grover circuits, thereby obtaining the iteration numbers along with the corresponding quantum oracles of the victim Grover circuit.


Prior research used power-based SCA for CO attack~\cite{ccspscaquantum}. However, our findings are in contradiction with the previous outcomes of power-based SCAs. This inconsistency arises due to the inadequate number of measurements performed on Grover oracle in~\cite{ccspscaquantum}. \textcolor{black}{Their results only offer precision up to two decimal points and derive conclusions based on these limited measurements.} Furthermore, our timing SCA yields better results with more measurements, as most of the overlapping coefficients from the same number of iterations exceed 0.99. With a higher quantity of measurements, the distinctions become evident, allowing for the differentiation between the two different Grover oracles.




\subsubsection{Circuits Ansatz Identification (CA)}\label{sec:ca}
\begin{figure}[bt!]
\centering
\includegraphics[width=\columnwidth]{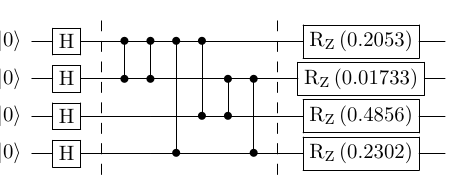}
\caption{QAOA circuit used to perform the timing SCA attack. The attack goal is to retrieve the 4 parameters in the $R_z$ gates.}
\vspace{-5mm}
\label{fig:qaoacircuit}
\end{figure}

\begin{figure*}[bt!]
\centering
\includegraphics[width=2.05\columnwidth]{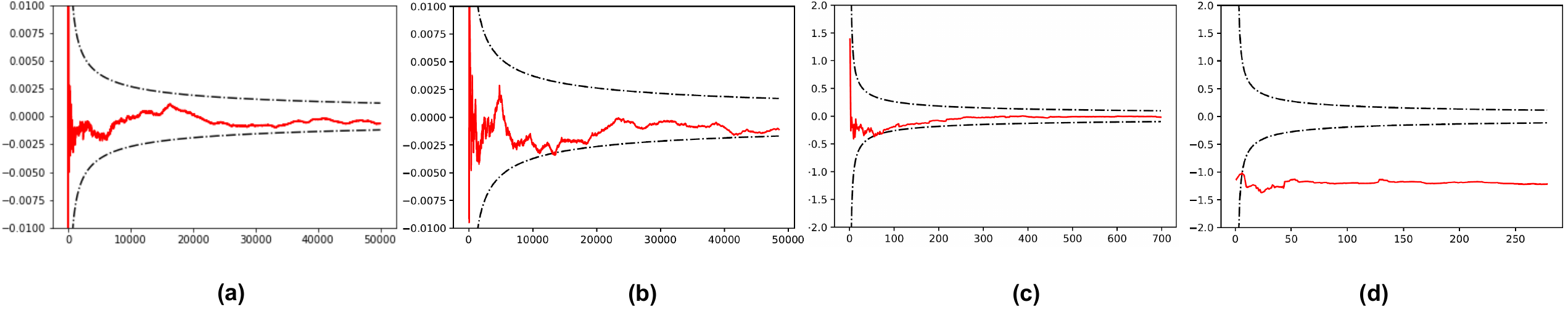}
\caption{DoM tests on two different parameters on 4 different datasets. \textit{x-axis} refers to the number of shots on two different datasets and \textit{y-axis} refers to the DoM between the two datasets in seconds. The grey line refers to the confidence interval of 95\% if two datasets have the same statistical states: (a) two the QAOA ansatz circuits with different parameters; (b) two different compilation strategies of \textit{HHL algorithm} circuit on a simulator; (c) 700 shots on two different compilation strategy of \textit{HHL algorithm} circuit on quantum computer \textit{ibm\_perth}, The DoM curve of (a), (b), and (c) are in between the two confidence interval, meaning that the two datasets cannot be distinguished statically. (d) DoM tests of 270 shots on an identical Grover search circuit using both the \textit{ibm\_perth} and \textit{ibm\_lagos} devices. With only 10 measurements, the DoM exceeds the confidence interval, indicating that the two datasets are statistically distinct from each other.}
\label{fig:domtests}
\end{figure*}

The Quantum Approximate Optimization Algorithm (QAOA) is a quantum computing technique used for solving combinatorial optimization problems efficiently~\cite{qaoa}. It requires computation from both quantum computers and classical computers altogether. In this paper, we selected a 4-qubit QAOA ansatz circuit with four parameters built in the quantum circuit as shown in Figure~\ref{fig:qaoacircuit} to perform the timing SCA attack. 
We performed the timing SCA on a simulator-based setup and we observed that the quantum simulator consumes the same amount of time to simulate identical quantum circuits with different parameters. This is because the quantum simulator employs a matrix representation and performs matrix multiplications each time a new gate is added to the quantum circuit.
Consequently, the total computational effort remains constant for a specific quantum circuit, resulting in an absence of significant time variations with different parameters of the $R_z$ gates. To illustrate this point, Figure~\ref{fig:domtests} (a) showcases an example of executing the QAOA ansatz with the same structure but distinct parameters on a simulator. Even in a real quantum computer execution, different parameters of $R_z$ gate do not alter the time difference of the quantum gate execution, since different parameters of the $R_z$ gates only change the pulse amplitude, which does not affect the time traces of the entire circuit execution. 

As seen from prior research~\cite{ccspscaquantum}, even the power SCA cannot retrieve the parameters of the quantum ansatz. This is because only the rotation angle of the $R_z$ gates in the ansatz circuit are different. Therefore, the CA attack is secure from both timing and power SCA attacks on local and cloud-based quantum computing services.

\begin{figure}[bt!]
\centering
\includegraphics[width=\columnwidth]{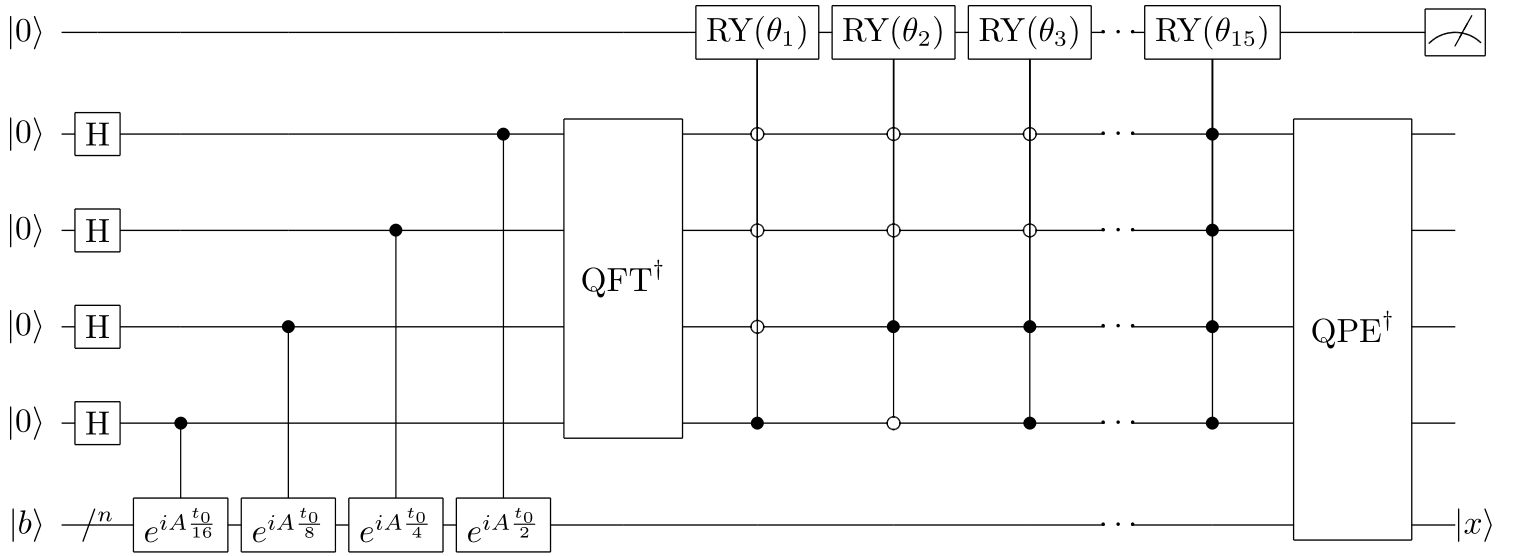}
\caption{General \textit{HHL} circuit. Multiple multi-qubit gates are used in the quantum circuit to magnify the difference in different routing techniques.}
\vspace{-5mm}
\label{fig:hhl}
\end{figure}
\subsubsection{Qubit Mapping Identification (QM)}\label{sec:qm}

In this subsection, we delve into the Qubit Mapping Identification process by conducting an attack on the targeted quantum computer. Specifically, we utilize the \textit{HHL} circuit to demonstrate this attack~\cite{harrow2009quantum}. An \textit{HHL} circuit is a linear matrix solver that solves the problem $Ax = b$ where $A$ is a matrix, $b$ is a vector, and $x$ is another vector that is being solved. A general \textit{HHL} circuit is demonstrated in Figure~\ref{fig:hhl}. We utilize the HHL circuit as the attack target because there are many multi-qubit gates such that many SWAP gates are required to perform the computation. We select two different compilation strategies to make sure they do not generate the same compilation result.

We conducted a timing SCA on the quantum computer using two different routing methods: SABRE and stochastic routing. The SABRE technique is a routing approach that helps address the routing problem on a noisy quantum computer~\cite{sabre}. The stochastic routing technique was developed by the Qiskit team and generates routing results that are slightly different from the SABRE technique's results. \textcolor{black}{The DoM tests on the two different routing techniques with identical HHL 
circuits are illustrated in Figure~\ref{fig:domtests} (b).} The nature of timing traces of these compiled circuits' execution suggests a slight variance in their expected time consumption. Surprisingly, after subjecting the \textit{HHL} circuit to 10,000 iterations using the two distinct routing methods, the execution times show minimal and hardly discernible contrast.

This outcome can be attributed to the limited modifications introduced to the two quantum circuits through the utilization of different routing methods. Additionally, for smaller circuits, modernized routing techniques generally excel at generating near-optimal solutions at a global level. This implies that varying routing methodologies tend to produce indistinguishable routing outcomes. Therefore, it becomes evident that the timing SCA fails to compromise the integrity of the Qubit Mapping technique. The intricate interplay between circuit size, routing efficacy, and the subtle divergences introduced by routing methods collectively contribute to the resilience of the Qubit Mapping technique against timing SCA.

Furthermore, we also performed the QM attack on the quantum computer \textit{ibm\_perth}. We changed the initial mapping on the compiled \textit{HHL} circuit and performed 700 measurements, as shown in Figure~\ref{fig:domtests} (c).
Since we observe that there is little to no difference between the execution of different quantum circuits, the QM is likely to be secure from timing SCA. Similarly, the similarity of different compilation strategies or differences in the initial mapping of qubits makes it difficult to backtrack the original qubit mapping of the quantum circuit using only timing information. Previous research has measured the power consumption of quantum computations and observed differences in power consumption when changing the order of qubit mapping, allowing the authors to distinguish between different initial mappings when at least one qubit differs in the initial layout. However, for the timing SCA, we did not observe any timing differences based on the initial mapping and compilation differences. 
We argue that the difference is shown in the per-channel computation of the quantum circuit execution since it is detected by the power-based SCA, but the cumulative computation eliminated the minor difference and is hard to distinguish the difference statistically.
Therefore, the proposed timing SCA does not support the QM attack.



\subsubsection{Quantum Processor Identification (QP)}\label{sec:qp}

The Quantum Processor timing SCA (QP) method enables the extraction of distinct features from quantum processors by examining the temporal overheads associated with executing identical quantum circuits on disparate quantum computers. Through this technique, various quantum chips, each characterized by distinct operational frequencies and time latencies, can be efficiently discerned by the QP using a modest number of measurements. This phenomenon is exemplified in Figure~\ref{fig:domtests} (d), which portrays the temporal consumption of a uniform quantum circuit on the \textit{ibm\_perth} and \textit{ibmq\_lagos} devices. The graphical representation reveals substantial disparities in the temporal consumption between the two quantum computers. Despite their shared attributes of possessing an identical number of qubits and architectural design, the execution times of the quantum circuits markedly diverge. The underlying distinction in execution times becomes especially evident through the DoM test depicted in Figure~\ref{fig:domtests} (d). This test effectively illustrates that, by executing merely five measurements on the quantum circuits, the QP method promptly and accurately distinguishes the specific quantum computer being targeted. 

Prior research also performed the power SCA on 9 different quantum circuits and concluded that it is easy to distinguish different individual circuits easily. Our experiments also agree with the power-based SCAs to perform the QP easily.


\subsection{Potential Strategies to Mitigate Timing SCAs}
In this section, we will discuss the possible defense that might be robust to the timing-based SCAs.

\subsubsection{Import randomness in the compilation process}
By importing randomness in the compilation process by changing the qubit index or executing the quantum circuit at the different locations of the quantum chip, the computation time can be altered. For instance, a 5-qubit quantum circuit can be executed at any 5 connected qubits in a quantum computer, with two different executions. The quantum chip routing map of \textit{ibm\_belem} is shown in Figure~\ref{fig:gates} (e). Therefore, different compilation strategies and time consumption do not affect the result of the computation. With more variation of the time consumption, it is more difficult for the attacker to discover the types of compilation and certain quantum circuits.

\subsubsection{Import spacial multi-programming system}
Spacial multi-programming systems are widely discussed and optimized for maximizing quantum resources while maintaining optimal fidelity and efficiency. When multiple jobs are submitted to the cloud systems, the quantum computer scheduling system can potentially alter the sequence of the jobs to optimize the qubit usage by allowing multiple quantum circuit execution concurrently. If the queuing system is designed to optimize the usage of quantum computers by observing the qubit number of the pending circuit, the attack method by submitting the quantum circuits during the intersection of the victim circuit might not be feasible anymore. However, the multi-programming system might cause some other security problems, as mentioned in Section~\ref{sec:related}. 

\subsubsection{Modify quantum circuits}
Minor modifications to the quantum circuit can enhance the security level of the quantum circuits. This observation stems from the fact that even slight adjustments to the quantum oracle can be detected by timing-based SCA. Introducing minor changes to the quantum circuits can potentially lead to significant alterations in the timing signature, rendering the quantum circuit more resilient against SCAs. For instance, inserting an identical gate into the quantum circuit can effectively alter the time consumption of each quantum circuit execution. Consequently, the overlapping coefficient of the same benchmark circuit is lower than anticipated, potentially bringing it closer to other baseline benchmark circuits.

\subsubsection{Add noise to timers and enforce privilege access}
To enhance protection against timing SCAs, one effective method involves adding controlled noise to timers and enforcing privileged access to specific users. Similar approaches are also undertaken in classical cloud service to protect from timing attacks. This involves quantum cloud providers incorporating noise into the timing mechanisms of their quantum processes. This extra noise acts as a defense against potential timing SCAs in the quantum system, making the quantum cloud infrastructure more secure. Simultaneously, by ensuring only authorized users with proper permissions can use quantum resources, the security from timing-based attacks will be further strengthened. This combined approach not only safeguards sensitive quantum computations and prevents unauthorized access, but also enhances the overall reliability and integrity of the quantum cloud services. This creates a more resilient and trustworthy environment for quantum cloud services. 
\section{Conclusion}\label{sec:Conclusion}
Security cannot be an afterthought and it is important to invest in quantum computer security now rather than wait for the whole ecosystem to collapse (as we have seen in classical computing). In this paper, we propose timing-based SCAs on IBM's quantum cloud service that have the potential to uncover the victim's quantum circuit concerning the quantum oracle. We utilized the threat model that requires fewer assumptions compared to prior works. We evaluate five different types of attacks and conduct simulations executed on a real quantum computer for each timing SCA. With just around 10 measurements, we can pinpoint the specific quantum processor being utilized. Additionally, we can classify the type of the victim's quantum circuit with a minimum of 1 measurement and a maximum of 18,712 measurements. Furthermore, the Grover oracle circuit can be reconstructed with a minimum of 500 measurements and a maximum of 20 million measurements. Finally, we suggest several approaches that might safeguard against the timing SCAs using different methods. In the future, we intend to scale our attacks to other quantum computers and circuits.

\balance
\bibliographystyle{IEEEtran}
\bibliography{references}

\end{document}